# Rapid Exploratory Maps and their potential role in supporting evidence-to-policy dialogues and processes

Gareth J Hollands[1*] (gareth.hollands@ucl.ac.uk),
Ian Shemilt[1] (i.shemilt@ucl.ac.uk),
James Thomas[1] (james.thomas@ucl.ac.uk)

(1) EPPI Centre, UCL Social Research Institute, University College London, London, UK

* Author for correspondence

## Abstract

Producing useful evidence syntheses requires clearly formulated research questions linked to an understanding of their optimal scope. When policymakers (or other end users) commission or steer the production of systematic reviews or other evidence syntheses to fulfil a demand for evidence, these parameters are rarely specified sufficiently for production to proceed. Instead, these details are typically developed via a collaborative dialogic process involving evidence producers and users to ensure the resulting synthesis will meet users' needs. This process is informed by scoping and (more-or-less formally) mapping the existing evidence base, which can take many weeks or months, slowed by applying mostly manual processes to identify and characterise the evidence. Recent developments in artificial intelligence (AI) technologies have opened up new opportunities for more rapid, iterative, and responsive mapping of relevant evidence to support such early-stage evidence-to-policy dialogues and processes. In this article, we outline and characterise this emerging approach, termed Rapid Exploratory Maps (REMs). We also present two illustrative case studies drawn from first-hand experiences within a programme of evidence reviews production to inform government policy in England. While this approach could be transformative for the mapping process, its broader potential is as yet largely unassessed, and it confers important challenges and likely risks, as well as benefits. Determining whether REMs could usefully support more efficient ways of working with end users when developing and designing evidence syntheses will require continued development and careful evaluation.

## Introduction

Policy decisions are not always informed by the best available evidence, reflecting the complex, non-linear nature of evidence-policy processes[1-4]. The likelihood of evidence being used in decision-making is, however, increased by timely access to relevant research findings, and relatedly, effective collaboration and relationships between researchers and policymakers[5]. This includes policymakers (or other end users) directly commissioning and/or steering the production of evidence syntheses (e.g. systematic reviews) to inform the development of policies, programmes, interventions or similar. However, producing useful evidence synthesis requires clearly formulated research questions linked to an understanding of the optimal scope, which determine the principal criteria used to assess evidence for eligibility, and selection of appropriate methods. At the outset, these parameters are rarely specified sufficiently for the researchers and policymakers to fully understand the nature and implications of the request, and to enable production to proceed. Instead, these essential details are typically developed via dialogue between evidence producers and end users to ensure the resulting synthesis will meet users' needs.

Formulating clear and viable research questions and informing design considerations for subsequent investigation, typically requires researchers first to scope, and to some degree to organise or map, the existing evidence base to appraise its nature and extent (including apparent gaps). This scoping process informs decisions about which bodies of evidence will be targeted and how, balancing the likelihood of generating robust and useful findings with feasibility[6-7]. Reviews started without sufficient understanding of the nature and extent of the available evidence can risk generating unrealistic expectations in policymakers of, resulting in disappointment in the utility of the review (and potentially a loss of confidence in using research in the future). However, a thorough scoping process, which may involve formal systematic evidence mapping processes,[7] can take many weeks or months, requiring mostly manual processes to identify and characterise the evidence. This can also be quite an inflexible process, which does not easily adapt to changing requirements as this process unfolds (e.g. changes in research questions or key eligibility criteria resulting from shifts in policy priorities or shared understandings about what might be possible or useful).

Recent developments in artificial intelligence[1] (AI) technologies, including the latest large language model (LLM)[8-9] based tools, alongside longer-established machine learning tools for classifying documents[10-11], have opened up new, previously impracticable opportunities for more rapid, iterative, and responsive mapping of relevant evidence. Here we outline this approach, which for brevity we term Rapid Exploratory Maps (REMs), and the implications for a possible role in supporting evidence-policy dialogues and processes. While this could be transformative and thus merits attention, it is as yet largely unassessed and confers important challenges and potential risks as well as benefits.

As this approach is newly emerging, the primary aim of this article is to introduce and explore this concept and illustrate how it has and could be applied in practice. It is informed primarily by the authors' first-hand experiences of developing evidence maps, including early experiences of developing rapid AI-enabled variants for discussion with policy teams, within a long-term programme of evidence reviews production to inform healthcare policy in England[2]. It is therefore not intended to provide detailed methodological guidance, as there is no single prescribed process or set of established standards known to maximise effectiveness. Accordingly, we recognise that these methods, the range and nature of their application and related conceptualisation will evolve with further, wider use.

---

[1] The capability of computational systems to perform tasks associated with human intelligence.

**Rapid Exploratory Maps (REMs) and their characteristics**

Evidence maps are systematic visual and interactive representations of the characteristics of available evidence relevant to a research question[12-13], and which commonly function as standalone products to be made publicly available for widespread use. However, because they provide users with an overview of the distribution of evidence within a broad research landscape (as well as the possibility to focus on individual records), they are also used to inform early-stage dialogue with interest-holders to help guide and design subsequent research[14-17]. Despite their different use cases, the latter use does not necessarily take less time or effort than the former,.

By more rapidly expediting evidence maps in evidence-policy contexts, REMs provide a tool to enable research questions, and key parameters for subsequent research (e.g. systematic reviews or evidence maps, and in principle, primary research studies), to be discussed, honed, and provisionally agreed much faster than has hitherto been the case. In short, they can help determine what end users are most interested in, whether this focus is likely viable and justifiable, and suggest the most appropriate methods to apply. As such, they reduce the time and resources that need to be applied to initial scoping and mapping, to a degree appropriate to their intended purpose. We summarise the key characteristics of REMs in Box 1, considering their combination as typical and relatively novel. This description is supplemented by Box 2, in which we present a possible process for their production, drawing on our prior experiences of developing them (see also Table 1 for case studies).

We do not consider REMs to be a distinct method per se, and they share commonalities with the constellation of established types of evidence mapping and scoping . This includes similarities to rapid variants of evidence maps previously characterised in the literature, including rapid Evidence and Gap Maps (EGMs)[18]. Both represent visual means of representing research activity, produced in short timeframes. Data extraction (i.e. classifying the characteristics of research studies into categories) may operate at a high-level, applying pre-existing coding schema alongside developing and refining new categories; and both will feature a relatively open initial scope, using literature searches limited in their numbers of sources but with broad terms and high yields. They will potentially include very large numbers of studies, and any analysis will likely be limited to the summarisation of high-level coding data (e.g. counts of different categorisations).

---

[2] Funded by the Department of Health and Social Care since 1995, and currently named the NIHR London-York PRP Evidence Review Facility.

**Box 1 – Key characteristics of Rapid Exploratory Maps (REMs)**

1) **Inform dialogue between researchers and interest-holders to specify an evidence request**
The primary overarching characteristic of REMs is the function they are intended to fulfil, namely supporting pragmatic and responsive early-stage dialogue between researchers and commissioners or other end users (e.g. policymakers) to formulate appropriate research questions and address related design considerations. Commonly we expect this to occur in the context of a time-sensitive policy demand or commission to produce a systematic review or other evidence synthesis. This primary identifying characteristic largely dictates the further secondary characteristics 2), 3) and 4) (below) which will typically arise as they reflect what is appropriate for, and beneficial to, the intended function.

2) **Very rapidly produced**
They are produced within a very short timeframe (e.g. weeks or days), which would reduce the resources associated with longer production cycles. Such rapid production processes also enable the scope and parameters of the map to be iterated responsively with back-and-forth interaction with end users, including the possibility that aspects of the map's parameters could be examined or adjusted dynamically within joint meetings.

3) **Require AI tools**
Typically, such rapid production dictates a substantial degree of automation, including screening and data extraction (i.e. coding or classification) processes that are intensively supported by AI technologies. Critically, for REMs, increasing automation cannot be considered optional or as only a goal to work towards. Instead, to make their production feasible, AI tools will likely need to be deployed at an early stage in their production, following rapid evaluation of performance within that context.

4) **Relative prioritisation of speed over accuracy**
REMs are not inherently inaccurate (and indeed this will vary by e.g. the map's focus and coding schema, and its degree of manual curation, oversight and validation); rather, they aim to be sufficiently accurate to achieve their primary intended purpose. Furthermore, because REMs are a tool largely defined by their typical use rather than by being a distinct, new product, their characteristics can be adaptable (e.g. they may become more or less 'quick and dirty' as the process unfolds).

**Box 2 – Illustrative possible process for producing a Rapid Exploratory Map (REM).** *This is suggestive of likely key steps based on our early experiences (see Table 1 for example case studies), and neither definitive or prescriptive, particularly given that the evolving and pragmatic nature of both this approach and evidence-policy contexts in general often require responsive and iterative adjustments to such research processes*

**1) Researchers receive a broad evidence request that will require conducting evidence synthesis work.** This could be from commissioners who also fund the research or via a collaboration with other intended end users

**2) First joint meeting between researchers and end users to begin specifying the evidence request.** Discussion to identify and agree a set of broad and inclusive characteristics for what should be considered as potentially relevant evidence, including the topic focus and characteristics (e.g. using PICO), accompanied by any known exclusion criteria or firm boundaries. As a result, the researchers produce a summary document setting out these characteristics, being sufficiently widely specified to accommodate a range of possible topic foci and inclusion/exclusion criteria, which is shared for further asynchronous feedback to inform production (stage 3).

**3) Production of the REM. Key elements include:**

- **Broad literature searches.** Wide-ranging in scope (e.g. inclusive sets of search terms) but rapid (e.g. limited sources), and conducted by an experienced information specialist.
- **Rapid AI-assisted screening for inclusion/exclusion and coding (i.e. classification) of a small number of key topic foci and study characteristics.** Selection of parameters designed to allow assessment of viability in relation to the body of evidence meeting the relevant broad topic focus and criteria for characteristics of that evidence (e.g. population, type of research design) that are known to be or could be of interest. Initially, include/exclude screening decisions could be made by a single human researcher supplemented by discussion with a second researcher. Once a set of benchmark manually-screened records have been collated (e.g. >100 records), an LLM can be prompted to make include/exclude decisions on this same benchmark set of records, to corroborate adequate include/exclude screening performance relative to human decisions, prioritising recall (i.e. ensuring that the machine is conservative and does not substantially exclude relevant records). Providing adequate performance is reached (or an explicit decision is made that the degree of suboptimal performance represents an acceptable trade-off), LLM decisions can then be used to automate the remainder of screening. A similar approach could also be used for coding categories of broad topic foci and the most pertinent study characteristics needed to inform the principal design decisions.
- **Software tools.** Required for i) managing bibliographic data from searches and supporting the stages of evidence synthesis production, ii) for producing evidence maps, and iii) deploying generative AI to enable LLM-supported screening and coding decisions. For example for case study 1 (Table 1), we used respectively, EPPI Reviewer, EPPI Visualiser, and OpenAI GPT-4o (being directly accessible within EPPI Reviewer).

**4) Second joint meeting between researchers and commissioners or end users to present and discuss the REM and further specify the evidence request.** Researchers present the REM and related findings regarding viability and possible options for narrowing or expanding the scope, and any key remaining uncertainties about parameters of the design of the subsequent evidence synthesis. Discussion could draw on specific borderline and exemplar cases and check interest in related topic foci, and across range of research characteristics (e.g. study designs, populations). Relatedly, attempt to reach consensus on specific research questions and linked inclusion/exclusion criteria that have not previously been confirmed. A revised summary document outlining a proposal for the subsequent evidence synthesis work is produced, which is shared for further asynchronous feedback.

**5) Begin production of a protocol for subsequent evidence synthesis work.** This is dependent on reaching a sufficient degree of consensus between the researchers and the commissioner or end users on the research questions, the linked inclusion/exclusion criteria, and any related design considerations for the future evidence synthesis work. If there are still important uncertainties to address, additional synchronous and asynchronous feedback activities can be conducted.

However, the key differentiating aspect of REMs is that they will be principally used only within, and not shared for use beyond, early-stage dialogue to specify an evidence request for prospective research. This is not a requirement for other rapid evidence maps, which by contrast are typically conducted as means of *fulfilling* an evidence request and thus intended to be published, and made available for widespread use. Because we expect this early-stage dialogue to occur in the context of a time-sensitive policy demand or commissioning process, this primary overarching characteristic of REMs largely dictates their further secondary characteristics that are related to a high degree of rapidity and AI use. Furthermore, because they are unlikely to be available for wider use, they are also able to apply an appropriate balance of speed relative to accuracy and/or detail. Presuming both high speed and high accuracy are not achievable, this likely dictates some prioritisation of the former.

**Addressing potential risks**

Appropriate methodological choices could, as mentioned, include limiting the scope of the search strategy - which may result in an incomplete or biased representation of the evidence base - and automating screening and classification of records *en masse,* subject to possible scaled-up errors of AI tools. When deploying AI tools , we can set minimum acceptable benchmarks for accuracy relative to human benchmarks at levels appropriate to the context of use. Performance below these levels would necessitate further development of the automated process (e.g. refinement of LLM prompts), supplementary manual resource being applied and/or explicit discussion with users of the risks in proceeding. However, even if acceptable benchmarks are met for overall performance, this could mask poor underlying performance concerning certain sets of characteristics (e.g. relevant studies that use terminology from a different discipline, or types of grey literature with more limited abstracts). The resulting REM may therefore inadequately reflect the true distribution of the literature's characteristics and any faulty assumptions that are derived risk being compounded in designing and conducting future work that is directly informed by the REM.

These risks can be mitigated initially by, for example, aiming to corroborate the performance of automation tools on heterogenous bodies of records varying in key characteristics. Furthermore, as the REM is completed and interpreted, key assumptions likely to inform major design decisions for future work, such as those relating to the relative absence or preponderance of important types of literature can be further checked such as by additional targeted searches and/or manual verification of accuracy. Such additional corroboration could continue until uncertainties of all parties were sufficiently assuaged, but naturally this would require additional time and resource. Because the development process is intentionally dialogic, it will benefit from as much oversight as well as engagement with the core literature as involved parties are able to provide.

Accordingly, initial decisions concerning pragmatic methodological trade-offs that could impact eventual accuracy and completeness, should be made in explicit consultation with end users, including presenting the potential benefits, risks and mitigations. Similar discussions should also accompany any interpretations or assumptions derived from the REM that is produced. Ultimately, if the commissioners or other users (or indeed the researchers themselves) do not understand the potential risks of inaccuracies in the REM and resulting unreliable and misrepresentative findings, they may place greater trust than is justifiable, which could shape decisions and future directions. These discussions between producers and users could usefully draw from published guidance for considering the appropriateness and balancing the inherent trade-offs for other rapid evidence synthesis products[18-19].

A further risk of rapid scoping and formulation of research questions in the context of a REM, is that this may mean less overall time is spent on thinking and discussion between researchers and interest-holders. While less time is not inevitably detrimental if it results in a more tightly focused process, engagement and feedback activities with interest-holders should be planned to mitigate this risk, whether synchronous (e.g. meetings between researchers and commissioners to present preliminary observations or discuss key decisions) and/or asynchronous (e.g. commenting on shared summary notes or iterations of protocol documents).

Further methodological development of REMs and related processes - including formal evaluations documenting procedural details and timings, and assessing a range of outcomes and perspectives - could support a deepened understanding of such potential risks and mitigations, including evidence of what can be gained and lost relative to existing approaches. Such development also includes emerging guidance about the responsible use of AI[20] (see Box 3 for a specific summary for this context), albeit the widespread and routine deployment of AI tools may remain contentious due to its societal and environmental impacts[21-22]. While acknowledging the various potential risks we have outlined, in the next section we instead consider the potential of REMs to be beneficial and useful.

**Box 3 – The responsible use of AI in REMs**

Compressing work that conventionally takes several months into weeks, days (or even less) using AI necessarily involves the acceptance of some risk, and the likelihood that accuracy is being traded for speed. Guidance for the Responsible use of AI in Evidence SynthEsis (RAISE)[20] provides assistance in navigating these competing priorities in accordance with principles of transparency and accountability. When conducting a REM, an evidence synthesist needs to explicitly justify their use of an AI tool and describe its strengths and weaknesses[20]. They accept that they are responsible for the claims made in the REM (not the AI tool or its developers) and for ensuring that the REM is used only for the purpose intended, and not disseminated further and portrayed as a comprehensive and authoritative view of the evidence base. If possible, it will be helpful if some indication of accuracy and comprehensiveness were given, though this requires more manual work in checking machine decisions manually. Where the accuracy of a REM has not been established as sufficiently high for decision-making purposes (either because manual checking has revealed weaknesses in AI classifications, or because there has not been time to undertake sufficient manual evaluation), the RAISE guidance states that it can be used for ‘exploratory and supplementary’ purposes only[27]. Where the REM is to be used as an aid to early-stage discussion and to help researchers and policymakers come to a joint understanding of where it may be helpful for an in-depth review to focus, this use case is likely to fall under ‘exploratory’ purposes and so be considered a responsible use of AI. If the distribution of research activity portrayed in the REM is more central to a decision (for example, when identifying genuine research gaps for commissioning new research), then unless sufficient accuracy has been established, this would fall outside ‘exploratory and supplementary’ use and not be considered an acceptable use of AI.

**How Rapid Exploratory Maps could be useful**

In Table 1, we summarise three specific ways in which REMs might be used to support evidence-policy processes, each accompanied by real-world examples drawn from two case studies. These examples demonstrate how such maps can function as a tool for progressing from initial broad requests for policy evidence, to the design of specific evidence synthesis products to address commissioner (and other user) needs. They reflect our direct early experiences in the context of national-level policymakers acting as the principal source of commissioning requests (and linked research funding) for evidence synthesis products, and as the principal end users of those products. Although this context has informed our expectations about the likely principal uses of REMs, in principle they could be relevant to a wider range of contexts and agents within broader evidence-policy ecosystems. This could extend to research funders or commissioners without any policymaking role, as well as policymakers or policy-adjacent organisations without any research funding or commissioning role but whom researchers may directly collaborate with to inform the development of useful evidence synthesis products. Approaches akin to REMs may also ultimately prove possible to employ in later phases of evidence-policy dialogues (such as being applied to supplement, or address queries arising from, already completed evidence syntheses); act as standalone products (potentially in conjunction with linked decision-making products[23]) when such rapid characterisation and appraisal of an evidence base is all that is required or possible; or act as a direct precursor to conducting primary research studies.

Producing REMs may also engender more general benefits beyond those outlined in Table 1. For example, the working relationship with commissioning interest-holders and others may benefit from having the map as a central visualisable focal point. This enables the research team to demonstrate that tangible progress is being made, as well as manage expectations early in the process. This could be by providing a clear indication if there is an apparent scarcity of evidence, or indeed by indicating that relevant evidence is so abundant or heterogenous as to suggest practical challenges to any subsequent evidence synthesis. Importantly, REMs are inherently collaborative, and can in principle be shaped responsively via repeated discussions and interactions with users, developing shared understanding and decisions. Similar to any co-production of research evidence, these processes could thus help to strengthen relationships and increase the likelihood that the resulting evidence synthesis answers end users' questions, caters effectively to their needs, and is used appropriately[24-25].

**Key technology enablers for now and the future**

Given technological developments have made REMs possible, in this final section we briefly describe their central role and consider how this may develop in future. Machine learning tools have assisted in evidence maps for some time[26]. For example, during the COVID-19 pandemic, two of the authors were involved in maintaining a progressively automated map of the evidence to assist the Department of Health and Social Care (England) in its pandemic response. The map began as a manual enterprise, with members of the research team

**Table 1 – Key potential uses of Rapid Exploratory Maps (REMs) to support evidence-policy processes.** Real-world examples are provided from two case studies involving the authors in the NIHR London-York PRP Evidence Review Facility. These case studies are intended primarily to illustrate the concept, being the result of retrospective reflection. Prospective formal evaluations based on detailed procedural data and assessing a range of outcomes will be needed to generate robust evidence on the relative benefits and limitations of these approaches.

**Case study 1: Health effects of e-cigarettes (vapes)**

The Department of Health and Social Care in England (DHSC) commissioned the Evidence Review Facility to produce a living evidence map to be updated for five years and focusing broadly on the health impacts of e-cigarette use to support ongoing national-level policy developments including the Tobacco and Vapes Bill. To inform initial discussions about the specific scope of the evidence map and the research questions it should address, we produced a REM. While this included a focus on literature concerning potential harms and health effects linked to vaping (as this was the DHSC's primary interest at outset), it was also clear there were several cognate areas of potential interest, and so the REM had a considerably broader scope that also extended to, for example, trends in product use, and interventions. Following presentation and discussion of this REM with the policy team, consensus decisions were reached about scope, research questions and related inclusion criteria. These decisions directly informed production of a full living protocol for the living evidence map[28], which is now being produced.

**Case study 2: Artificial intelligence (AI) for administrative tasks in primary care**

DHSC asked the Evidence Review Facility to identify evidence on the use of AI in primary care. This broad request was made in the context of the NHS Long Term Plan[29], which proposed a shift from 'analogue to digital', including greater use of AI for administrative and clerical tasks in primary care to increase staff productivity and time to see patients. An initial REM of the broad literature revealed a substantially larger landscape of potential policy interest than was envisaged by the commissioners at the time of the initial request, particularly in terms of the wide range of AI tools and their applications to primary care tasks. This informed the scope for a subsequent systematic evidence map which in turn will help identify potential foci for in-depth systematic reviews.

| Potential use | Explanation | Case study examples, including reflections on perceived added value |
|---|---|---|
| **i) Rapid assessment of the viability of the general scope and broad research questions of commissioned** | Commonly when an evidence synthesis is first commissioned this consists of relatively high-level evidence requests. This may manifest in a general initial scope and broad research questions that are insufficiently specified, bounded or granular to usefully inform the design and production of subsequent reviews or maps by the commissioned research team. This is not a result of commissioners and users (e.g. policymakers) not knowing what they would like answers to, but instead typically reflects common differences in language and skills, as well as differing purposes | <u>Case study 1:</u> The initial evidence request primarily focused on health impacts of e-cigarette use. It was able to be determined that this was a viable literature to map with initial broad searches identifying ~14,000 MEDLINE records and semi-automated coding of records within the REM concerning health effects of vaping identifying >1200 records. Commissioners also expressed a possible interest in a potentially broader topic scope e.g. including studies on trends in use of vaping products, and on related policies and interventions to alter |
| **evidence syntheses** | and interpretations of and for the work. Through direct access to and dialogue between researchers and policymakers, evidence requests can be translated in order to reach a mutual understanding and formulate an appropriate scope and set of broad research questions.<br><br>In supporting this process, REMs can feature wide external boundaries that are able to accommodate identification of a range of possible broad research questions, extend across disciplines and literatures as policy questions frequently do, and include a small number of top-level categorisations of the topics and types of included evidence. This allows estimation and appraisal of the scale of the relevant evidence base, including whether it is sufficiently large to be usefully interrogated, or indeed may be too large to be practicable. Relatively rapid and responsive assessment of the general scope and viability, including the ability to approximate the effect of expanding or reducing the scope, can also aid management of the commissioners' expectations early in the process. For example, this could enable researchers to provide a clear indication that there is an absence or scarcity of evidence (or indeed that the proposed focus may be overly ambitious or overwhelming relative to the time available and will require further specification). | use, and these topics were also determined to be viable with >1000 relevant records identified within the REM. We perceived that the ability to rapidly categorise the broad topic focus of the large evidence base allowed commissioners to affirm their interest in these topics and to consider some expansions of their initial suggested scope. Furthermore, more detailed semi-automated coding of the types of linked health effects addressed by the literature revealed that a very wide range of health effects had already been examined beyond only impacts of vaping on tobacco use and cessation, which reinforced the value to the commissioners of being able to broadly monitor this evolving literature. Producing the REM allowed such initial quantified assessments of the breadth and depth of the literature, which would not have been possible at such scale if examined manually.<br><br>Case study 2: The initial interest of the policy team was on use of AI for non-clinical administrative tasks, but rapid mapping also encompassed clinical tasks and determined that the evidence base for those clinical applications was far larger (approximately ≥10 times) as that for non-clinical administrative tasks. This enabled a wider range of viable broad research questions to be identified and also increased policymakers' awareness of important related literatures that they were less aware of at the time of the initial request. This helped policymakers to discuss internally and check and reaffirm their continued specific interest in non-clinical administrative tasks. While a traditional scoping approach would likely have similarly identified the presence of a substantially larger evidence base for clinical applications, we perceive that the ability to estimate the actual relative magnitude of these literatures provoked greater consideration and discussion of these cognate areas even if they were ultimately not pursued, helping clarify the priority focus. |
| **ii) Focal point for detailed discussion to refine specific research questions and linked inclusion/ exclusion criteria** | Providing the broad viability of the commissioned evidence synthesis is ascertained, more detailed coding of the characteristics of included evidence can be developed and discussed. This could concern eligibility criteria (e.g. PICO) domains that are not already specified, supporting discussion of which criteria are important and how applying them may impact on any subsequent evidence synthesis. This could aid formulation of inclusion and exclusion criteria and refinement of more specific research questions. For example, end users may be unsure about some criteria (e.g. which study designs should be included or excluded). Coding a range of study design categories (e.g. randomised controlled trials, longitudinal studies, and so forth) would give an initial indication of the nature and extent of evidence that could be gained or lost by including or excluding particular designs. By contrast, users may already have a clear sense of other criteria (e.g. specific populations or interventions) but categorising data to encompass more detailed or broader ranges of related categories may reveal an extant evidence base in cognate areas that could result in re-appraising the optimal scope.<br><br>The discursive dialogic processes necessary at this stage may benefit from being able to visually represent and view the evidence and some of its possible categorisations, as well as examine sets of individual records with particular combination of characteristics within the map, including exemplars and borderline cases. The likely relatively rapid production cycle for REMs could enable the parameters, structure and presentation of the map to be iterated responsively with back-and-forth interaction with end users. | <u>Case study 1:</u> First, the process of developing automated coding tools and identifying borderline cases that arose led to discussions and clarifications concerning different types of eligible nicotine products and their relationship with use of other substances. This resulted in broadening the set of eligible products beyond e-cigarettes to also include other non-tobacco oral nicotine products such as nicotine pouches. , Second, borderline examples of in vitro studies arose when mapping types of health effects, which led to clarification that such studies should definitively be excluded. While a traditional scoping approach would likely similarly identify useful borderline cases for discussion, it would be less able to help rapidly identify sets of records that have been coded with particular characteristics to aid in these discussions.<br><br><u>Case study 2:</u> The challenge of attempting to specify inclusion criteria for the wide range of AI tools used in non-clinical administrative tasks (including generative and traditional AI tools), led to rapid development of a general typology of AI tools that may be relevant and within the broad scope. This informed detailed discussions to refine eligibility criteria e.g. decisions to include AI tools used to generate or improve personalised information given to individual patients by clinicians, but exclude AI tools used to generate generic materials developed for groups of patients. |
| **iii) Aiding in selection of appropriate research** | The production of REMs and the accompanying collaborative process could also inform changes in the evidence synthesis product that is ultimately requested by the commissioner or end user. This could plausibly be at the level of decisions about the research design itself, in | <u>Case study 1:</u> Commissioners initially requested a long-term living evidence map intended primarily for evidence surveillance and regular descriptive reports of its contents. The scale of the requested evidence synthesis product as well as its nature, suggested it would |
| **designs and methods, including via direct testing** | conjunction with following guidance about which may be most suitable, such as a systematic review (e.g.[30]), an umbrella review, or a living or static evidence map[13]. For example, should a REM identify marked gaps in key areas of policy interest this may be sufficient to advise against conducting an in-depth systematic review. Alternatively, the commissioner may have requested an evidence map to help identify areas that may merit future systematic reviews, but in developing an initial REM substantive clusters of evidence may be identified that may make a strong case for immediate development of a systematic review(s) instead. Alternatively, a number of existing systematic reviews could be identified thus suggesting against generation of another and instead supporting capturing these within a broader umbrella review or evidence map.<br><br>A REM may also inform decisions about specific methodological aspects within the design. For example, it could suggest provisional eligibility criteria, coding schema, and search strategies (including the ability to rapidly source reference records for calibrating searches). If an evidence map is the product that is to be subsequently produced, development of a REM may constitute a direct prototype and model for that. Furthermore, it could provide a testbed for assessing methodological options, in particular processes for large-scale AI-enabled automation of screening and classification (which may already have required some development and deployment in producing the REM). The opportunity to assess such methodological approaches can also contribute to formal and larger-scale methods evaluations, including of AI technologies for screening and data extraction in evidence synthesis[31]. | be an appropriate candidate for a substantial degree of automation. Automated screening and coding processes were developed and assessed within the REM, and performed sufficiently well to indicate that they could plausibly be used for the subsequent commissioned living map. As a result this semi-automated approach was embedded within the protocol[28] approved by the commissioners. These informative insights into likely viable screening and coding processes for the subsequent living map could not have been gained using a traditional manual approach, and so without the REM may have necessitated additional development work at a later stage instead.<br><br>Case study 2: The REM that was produced suggested an evidence base that was large, complex and with some key parameters not being well understood (e.g. the range of potentially relevant AI tools that may be encountered). Accordingly, this informed the decision that the next phase of the commissioned work should be a full systematic evidence map. This map aims to provide a detailed understanding of the relevant landscape, which will in turn allow the identification of potential foci for subsequent in-depth systematic reviews. We perceive that production and discussion of the REM aided in highlighting the wide range of AI tools examined in the literature and the related challenge of specifying inclusion/exclusion criteria, supporting subsequent production of a systematic evidence map rather than a narrower or more definitive evidence synthesis product (e.g. a systematic review). |

screening records for relevance, and classifying eligible records using a ‘keywording’ tool. This manual effort produced sufficient training data to train machine learning models, which enabled the team to automate much of the maintenance of the map. Critical here was the quantity of high-quality manually ‘labelled’ data from the map itself, which enabled accurate and specific machine learning tools to be built. The technologies available at the time required these training data to work, and automation would have been impossible without them.

The critical development in more recent machine learning technology, and the reason that REMs are possible, are ‘generative’ (or ‘autoregressive’) Large Language Models (LLM). This type of model is pre-trained, often on vast quantities of data, and can be used to assess map eligibility, or to apply ‘keywords’, without the need for map-specific training data. Thus, a language model can be presented with the results of a search for research and classify records according to their likely eligibility for a map, without needing any information beyond a clear definition of inclusion / exclusion criteria. Likewise, the model can also classify records according to, for example, their research design, study population, intervention and possible risks of bias. Even more recently, ‘decision-oriented’ language models, which are also able to classify text, have begun to appear. These may be even more suitable for REMs, as they are more consistent in their results, and can run much more quickly (and cheaply).

Following on rapidly from the emergence of LLMs as the dominant form of machine learning has been ‘agentic AI’. These are systems that can operate more autonomously than machine learning has hitherto done, and plan, make decisions, and take actions as part of a continuous workflow, rather than each step being designed and activated by a human. Agentic mapping software is now emerging that will: take a query from a user in plain English (i.e. not a structured search or even necessarily a PICO question); create and execute a search strategy; generate and apply screening criteria; design and apply a keywording tool; and provide an interactive visualisation of results. Figure 1 shows the results of such an agentic mapping process from work conducted in the Wellcome Trust-funded ‘DESTINY’ project (https://destiny-evidence.org/). Should robust evaluations of this approach find it to be sufficiently accurate, the time involved in generating useable REMs in future could be reduced to minutes.

**Conclusion**

Rapid Exploratory Maps (REMs) represent an early example of how generative AI tools, coupled with more established machine learning tools, are starting to be deployed within evidence synthesis production workflows. While they could prove to be transformative for the mapping process, their broader potential is as yet largely unassessed, and further methodological development and careful evaluation of potential beneficial and detrimental impacts, is needed. This could support their judicious application to facilitate evidence-policy processes and relationships, and the uptake of evidence.

**Figure 1: the results of a 10-stage agentic mapping workflow, from user query through to visualising results**

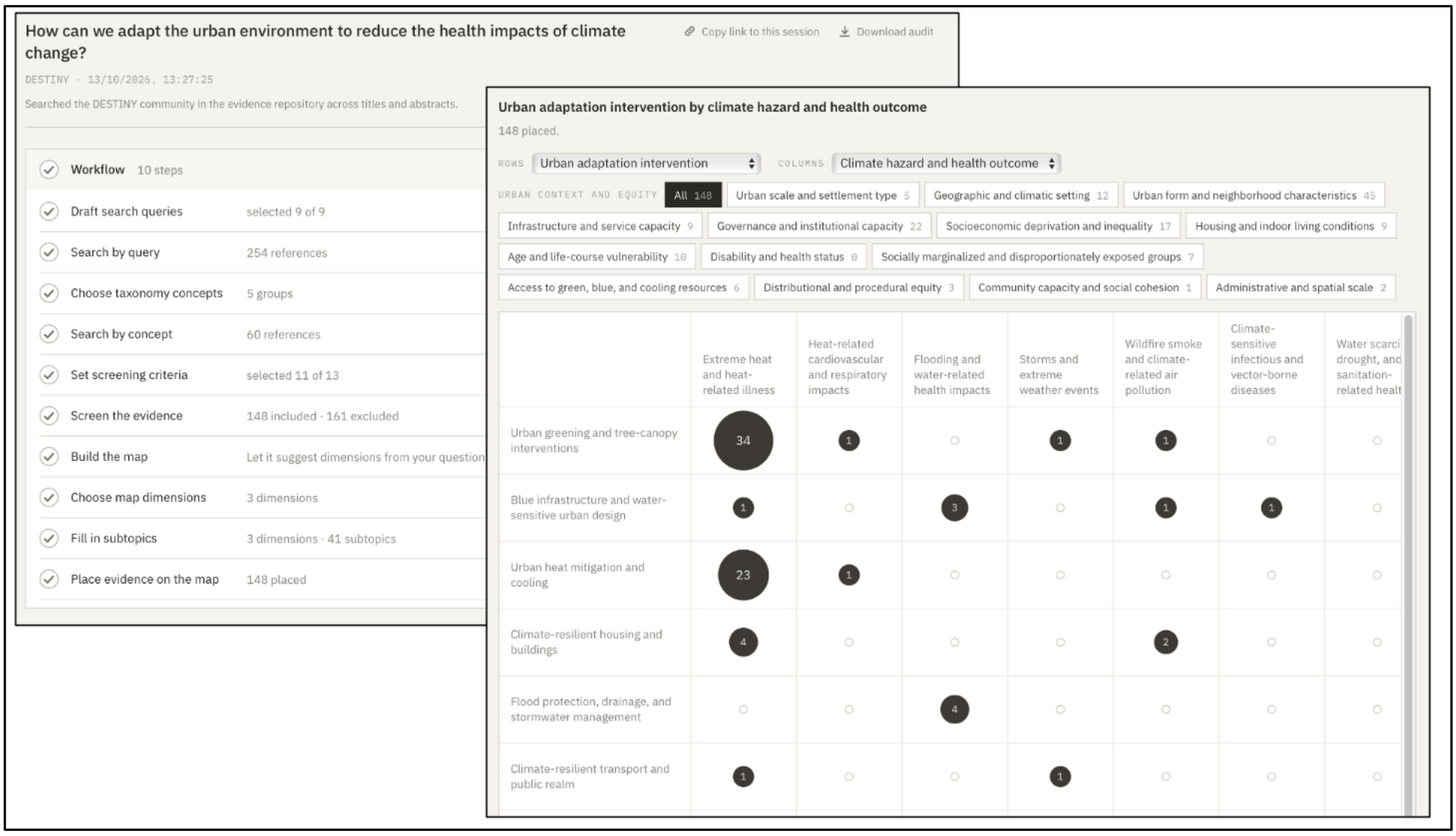

## References

1. Macleod MR, Michie S, Roberts I, Dirnagl U, Chalmers I, Ioannidis JPA, et al. Biomedical research: increasing value, reducing waste. The Lancet, 2014;383(9912):101-4. 10.1016/S0140-6736(13)62329-6
2. Langer L, Tripney J, Gough D. The Science of Using Science: Researching the Use of Research Evidence in Decision-Making. London: EPPI-Centre, Social Science Research Unit, UCL Institute of Education, University College London; 2016.
3. Miranda JJ, Beran D, Nugent R, Diez-Canseco F, Sargent JL, Madise N, et al. Maximizing researcher–policymaker engagement in global public health. Nature Medicine, 2025;31(11):3624-33. 10.1038/s41591-025-04015-9
4. Rutter H, Wabnitz K, Nambiar D, Garde A, Benton T, Heymann D, et al. The Lancet Commission on improving population health post COVID-19. The Lancet, 2026;407(10525):267–308. https://doi.org/10.1016/S0140-6736(25)02061-6
5. Oliver K, Innvar S, Lorenc T, Woodman J, Thomas J. A systematic review of barriers to and facilitators of the use of evidence by policymakers. BMC Health Services Research, 2014;14(1):2. 10.1186/1472-6963-14-2
6. Thomas J, Kneale D, McKenzie JE, Brennan SE, Bhaumik S. Determining the scope of the review and the questions it will address. Cochrane Handbook for Systematic Reviews of Interventions, 2019. p. 13-31. https://doi.org/10.1002/9781119536604.ch2
7. Sutcliffe K, Stokes G, O'Mara A, Caird J, Hinds K, Bangpan M, et al. Paediatric medication error. A systematic review of the extent and nature of the problem in the UK and international interventions to address it. London: EPPI-Centre; 2014. Available from: https://eppi.ioe.ac.uk/cms/Publications/Systematicreviews/Paediatricmedicationerror/tabid/3466/Default.aspx?tabid=273

8. Clark J, Barton B, Albarqouni L, Byambasuren O, Jowsey T, Keogh J, et al. Generative artificial intelligence use in evidence synthesis: A systematic review. Research Synthesis Methods, 2025;16(4):601-19. 10.1017/rsm.2025.16
9. Siemens W, von Elm E, Binder H, Böhringer D, Eisele-Metzger A, Gartlehner G, et al. Opportunities, challenges and risks of using artificial intelligence for evidence synthesis. BMJ Evidence-Based Medicine, 2025:bmjebm-2024-113320. 10.1136/bmjebm-2024-113320
10. Cierco Jimenez R, Lee T, Rosillo N, Cordova R, Cree IA, Gonzalez A, et al. Machine learning computational tools to assist the performance of systematic reviews: A mapping review. BMC Med Res Methodol, 2022;22(1):322. 10.1186/s12874-022-01805-4
11. Shemilt I, Simon A, Hollands GJ, Marteau TM, Ogilvie D, O'Mara-Eves A, et al. Pinpointing needles in giant haystacks: use of text mining to reduce impractical screening workload in extremely large scoping reviews. Research Synthesis Methods, 2014;5(1):31-49. 10.1002/jrsm.1093
12. Campbell F, Tricco AC, Munn Z, Pollock D, Saran A, Sutton A, et al. Mapping reviews, scoping reviews, and evidence and gap maps (EGMs): the same but different— the "Big Picture" review family. Systematic Reviews, 2023;12(1):45. https://doi.org/10.1186/s13643-023-02178-5
13. Hollands GJ, Shemilt I, Thomas J, et al. Living evidence maps: an emerging form of evidence synthesis. in preparation.
14. Miake-Lye IM, Hempel S, Shanman R, Shekelle PG. What is an evidence map? A systematic review of published evidence maps and their definitions, methods, and products. Systematic Reviews, 2016;5(1):28. 10.1186/s13643-016-0204-x
15. Bonell C, Jamal F, Harden A, Wells H, Parry W, Fletcher A. Systematic review of the effects of schools and school environment interventions on health: evidence mapping and synthesis. Public Health Res, 2013;1(1). 10.3310/phr01010
16. Kwan I, Hollands GJ, Macdowall W, Burchett HED, D'Souza P, Richardson M, et al. Equity impacts of active travel interventions: A systematic evidence map of the extent and nature of evaluation research. F1000Research, 2026.
17. Hollands GJ, Macdowall W, Burchett HED, Kwan I, D'Souza P, Richardson M, et al. Effects of active travel interventions in the UK on disadvantaged population groups and their potential to reduce health inequalities: systematic review. Journal of Epidemiology and Community Health, 2026;80:663-73. 10.1136/jech-2026-225997
18. Campbell F, Sutton A, Pollock D, Garritty C, Tricco AC, Schmidt L, et al. Rapid reviews methods series: guidance on rapid scoping, mapping and evidence and gap map ('Big Picture Reviews'). BMJ Evidence-Based Medicine, 2025;30(4):268-77. 10.1136/bmjebm-2023-112389
19. Garritty C, Nussbaumer-Streit B, Hamel C, Devane D. Rapid reviews methods series: assessing the appropriateness of conducting a rapid review. BMJ Evidence-Based Medicine, 2025;30(1):55. 10.1136/bmjebm-2023-112722
20. Thomas J, Hair K, Noel-Storr A, Aromataris E, Askie L, Berg RC, et al. Responsible use of AI in evidence SynthEsis (RAISE 2026) 1: Living recommendations for practice. Research Synthesis Methods, in press.
21. Belcak P, Heinrich G, Diao S, Fu Y, Dong X, Muralidharan S, et al. Small Language Models are the Future of Agentic AI. arXiv preprint arXiv:250602153, 2025.
22. Hosseini M, Gao P, Vivas-Valencia C. A social-environmental impact perspective of generative artificial intelligence. Environmental Science and Ecotechnology, 2025;23:100520. https://doi.org/10.1016/j.ese.2024.100520
23. Mohan A, White H. Enhancing Policy Impact Through Knowledge Translation: The Role of Innovative Evidence Products. Campbell Systematic Reviews, 2025;21(4):e70084. https://doi.org/10.1002/cl2.70084
24. Duncan S, Oliver S. Motivations for engagement. Research for All, 2017;1(2).
25. Oliver K, Kothari A, Mays N. The dark side of coproduction: do the costs outweigh the benefits for health research? Health Research Policy and Systems, 2019;17(1):33. 10.1186/s12961-019-0432-3
26. Shemilt I, Arno A, Thomas J, Lorenc T, Khouja C, Raine G, et al. Cost-effectiveness of Microsoft Academic Graph with machine learning for automated study identification in a living map of coronavirus disease 2019 (COVID-19) research [version 2; peer review: 2 approved]. Wellcome Open Research, 2024;6:210. https://doi.org/10.12688/wellcomeopenres.17141.2
27. Thomas J, Hair K, Noel-Storr A, Aromataris E, Askie L, Berg RC, et al. Responsible use of AI in evidence SynthEsis (RAISE 2026) 3: Living guidance for selecting and using AI evidence synthesis tools. Research Synthesis Methods, in press.
28. Hollands GJ, Shemilt I, Sutcliffe K, Stansfield C, Sowden AJ, Thomas J. Harmful health effects of e-cigarettes (vapes) and other non-tobacco oral nicotine products, trends in their use, and interventions: living protocol for a living evidence map and linked research digests [version 1; peer review: awaiting peer review]. F1000Research, 2025;14(609). 10.12688/f1000research.165927.1
29. Department for Health and Social Care. Fit for the Future: The 10 Year Health Plan for England. London: DHSC; 2025. Available from: https://www.gov.uk/government/publications/10-year-health-plan-for-england-fit-for-the-future.
30. Munn Z, Peters MDJ, Stern C, Tufanaru C, McArthur A, Aromataris E. Systematic review or scoping review? Guidance for authors when choosing between a systematic or scoping review approach. BMC Medical Research Methodology, 2018;18(1):143. 10.1186/s12874-018-0611-x
31. Shemilt I, Thomas J, Churchill R, Finnerty Mutlu A, Graziosi S, Khouja C, et al. Generative artificial intelligence tools versus conventional screening by humans for selecting eligible study reports for evidence synthesis: a living study within a review (SWAR). 2025. Available from: https://osf.io/g7mkb/overview